\documentclass[conference]{IEEEtran}
\ifCLASSINFOpdf
\else
\fi
\usepackage{enumitem}
\setlist[description]{font=\normalfont\itshape\space}

\usepackage{pgfplots}
\usepgfplotslibrary{external,colormaps,groupplots,statistics}
\pgfplotsset{compat=1.8}
\usepackage{tikz}
\usepackage{listings}
\usetikzlibrary{arrows,shapes,positioning}
\usepackage{multirow}
\usepackage{graphicx}
\usepackage{microtype}
\usepackage{balance}
\usepackage{xcolor}
\usepackage{xspace}
\usepackage{booktabs}

\definecolor{codegreen}{rgb}{0.25,0.5,0.35}
\definecolor{codegray}{rgb}{0.5,0.5,0.5}
\definecolor{codepurple}{rgb}{0.6,0,0}
\definecolor{backcolour}{rgb}{0.95,0.95,0.92}
\definecolor{colorstring}{rgb}{0.5,0,0.35}	
\definecolor{rltred}{rgb}{0.5,0,0}
\definecolor{rltgreen}{rgb}{0,0.5,0}
\definecolor{rltblue}{rgb}{0,0,0.5}
\definecolor{DarkGreen}{rgb}{0.00,0.60,0.00}
\definecolor{ScarletRed}{rgb}{0.80,0.00,0.00}
\definecolor{blizzardblue}{rgb}{0.67, 0.9, 0.93}
\definecolor{green-yellow}{rgb}{0.68, 1.0, 0.18}
\definecolor{dkgreen}{rgb}{0,0.6,0}
\definecolor{gray}{rgb}{0.5,0.5,0.5}
\definecolor{mauve}{rgb}{0.58,0,0.82}
\definecolor{lightgrey}{rgb}{0.90,0.90,0.90}
\definecolor{grey}{gray}{0.75}
\definecolor{light-gray}{gray}{0.80}

\lstdefinestyle{mystyle}{
	backgroundcolor=\color{backcolour},   
	commentstyle=\color{codegreen},
	keywordstyle=\color{colorstring}\bfseries,
	numberstyle=\ttfamily\color{codegray},
	stringstyle=\color{codepurple},
            basicstyle={\scriptsize\ttfamily},
	breakatwhitespace=false,         
	breaklines=true,                 
	captionpos=b,                    
	keepspaces=true,                 
	numbers=none,                    
	showspaces=false,                
	showstringspaces=false,
	showtabs=false,                  
	tabsize=2
}
\usepackage{boxedminipage}
\newenvironment{result}%
{\smallskip
\noindent
\let\emph=\textbf
\begin{boxedminipage}{\columnwidth}\begin{center}\em}%
{\end{center}\end{boxedminipage}%
}

\newcommand{\evo}{{\sc EvoMaster}\xspace}

\begin{document}

\title{RESTful API Automated Test Case Generation}

\author{
\IEEEauthorblockN{Andrea Arcuri}
\IEEEauthorblockA{Westerdals Oslo ACT, Oslo, Norway\\
and SnT, University of Luxembourg, Luxembourg\\
Email: arcand@westerdals.no}
}


%


\maketitle

\begin{abstract}
Nowadays, web services play a major role in the development of enterprise applications.
Many such applications are now developed using a service-oriented architecture (SOA),
where microservices is one of its most popular kind.
A RESTful web service will provide data via an API over the network using HTTP, possibly
interacting with databases and other web services.
Testing a RESTful API poses challenges, as inputs/outputs are sequences of HTTP requests/responses to  a remote server.
Many approaches in the literature do black-box testing, as the tested API is a remote service whose code is not available.
In this paper, we consider testing from the point of view of the developers,
which do have full access to the code that they are writing.
Therefore, we propose a fully automated white-box testing approach, 
where test cases are automatically generated using an evolutionary algorithm.
Tests are rewarded based on code coverage and fault finding metrics.
We implemented our technique in a tool called \evo, which is open-source.
Experiments on two open-source, yet non-trivial RESTful services and an industrial one, do show that our novel technique did automatically find 38 real  bugs in those applications.
However, obtained code coverage is lower than the one achieved by the manually written test suites already existing in those services.
Research directions on how to further improve such approach are therefore discussed.

\end{abstract}


%
\IEEEpeerreviewmaketitle

\section{Introduction}

Service-Oriented Architectures (SOA), and in particular Microservice Architectures~\cite{newman2015building},
are the common practice when building enterprise applications.
The market value  for SOA services was \$5.7 billions in 2013, 
where the market leaders are companies like IBM, Microsoft, Oracle and SAP.
Such market is estimated to reach \$16 billions by 2020\footnote{http://www.radiantinsights.com/research/services-oriented-architecture-soa}.
Currently, REST~\cite{fielding2000architectural} is the most common way to build web services used in enterprise systems.

Besides being used internally in many enterprise applications, there are many RESTful web services available on the Internet.
Websites like 
\emph{ProgrammableWeb}\footnote{https://www.programmableweb.com/api-research} 
currently list more than 16 thousand Web APIs.
In the Java ecosystem, based on a 
survey~\footnote{http://www.infoworld.com/article/3153148/java/oracle-survey-java-ee-users-want-rest-http2.html} 
of 1700 engineers, better REST support (together with HTTP/2 support) was voted as the most desired feature in the next version of Java Enterprise Edition (at that time, JEE 8).
This is because, according to that survey, ``The current practice of cloud development in Java is largely based on REST and asynchrony''.

Testing web services, and in particular RESTful web services, does pose many challenges~\cite{canfora2009service,bozkurt2013testing}.
Different techniques have been proposed, especially to handle the complexity of service orchestration, and black-box testing of external services.
Most of the work so far has been concentrating on SOAP web services.
SOAP is a well defined protocol based on XML.
However, most enterprises nowadays are shifting to REST services, which usually employ JSON (JavaScript Object Notation) as data format for the message  payloads.
Furthermore, there is not much research on white-box testing of web services, as that requires having access to the source code of those services.

In this paper, we propose a novel approach that can automatically generate integration tests for RESTful web services.
Our technique has two main goals: maximising code coverage (e.g., statement coverage), and finding faults using the HTTP return statuses as an automated oracle.
We aim at testing RESTful services in isolation, which is the typical type of testing done directly by the engineers while developing those services.
To generate the tests, we employ an evolutionary algorithm, in particular a Genetic Algorithm using the Whole Test Suite approach~\cite{GoA_TSE12}.

We implemented a tool prototype called \evo, and carried out experiments on three different RESTful web services.
Two are open-source, whereas the third was provided by one of our industrial partners.
Those systems range from 2 to 10 thousand lines of code.
Results of our experiments show that our novel technique did automatically find 38 real faults in these systems.
However, code coverage results are relatively low compared to the coverage obtained by the existing, manually written tests.
This is due mainly to the presence of string constraints and interactions with databases and external web services. 
Further research will be needed to address these issues to improve performance even further.

In particular, this paper provides the following research and engineering contributions:
\begin{itemize}
\item We designed a novel technique that is able to generate effective tests cases for RESTful web services.
\item We propose a method to automatically analyse, export and exploit white-box information of these web services to improve the generation of  test data. 
\item We presented an empirical study on non-trivial software which shows that, even if our tool is in a early prototype stage, it can automatically find 38 real faults in those RESTful web services.
\item To enable replicability of our results, tool comparisons and reuse of our algorithm implementations, we released our tool prototype under the open-source LGPL license, and provided it on the public hosting repository 
GitHub\footnote{https://github.com/arcuri82/EvoMaster}.    
\end{itemize}

\section{Background}

\subsection{HTTP}

The Hypertext Transfer Protocol  (HTTP) is an application protocol for communications over a network. 
HTTP is the main protocol of communication on the World Wide Web.
The HTTP protocol is defined in a series of  Requests for Comments (RFC) documents maintained by Internet Engineering Task Force (IETF) and the World Wide Web Consortium (W3C), like for example 
RFC 7230\footnote{https://tools.ietf.org/html/rfc7230} 
and RFC 7231\footnote{https://tools.ietf.org/html/rfc7231}. 

An HTTP message is usually sent over TCP, and is composed of four main components:
\begin{description}[leftmargin=2em]
\item[Verb/Method:] the type of operation to do, like getting a specific web page.
\item[Resource path:] an identifier to specify on which resource the HTTP operation should be applied, like for example the path of an HTML document to get.
\item[Headers:] extra metadata, expressed as a list of key/value pairs. 
An example of metadata is the \emph{accept} header, which is used to specify the format (e.g., HTML, XML or JSON) in which the resource should be returned (a resource could be available in different formats).
\item[Body:] the payload of the message, like the HTML text of a web page that is returned as response to a get request. 
\end{description}

The HTTP protocol allows the following operations (i.e., verbs/methods) on the exposed resources:
\begin{description}[leftmargin=2em]
\item[GET:] the specified resource should be returned in the body part of the response.
\item[HEAD:] like GET, but the payload of the requested resource should not be returned. 
This is useful if one only needs to check if a resource exists, or if he just needs to get its headers.
\item[POST:] send data to the server, e.g., the text values in a web form. Often, this method is the one used to specify that a new resource should be created on the server.  
\item[DELETE:] delete the specified resource.
\item[PUT:] replace the specified resource with a new one, provided in the payload of the request.
\item[PATCH:] do a partial update on the given resource. This is in contrast with PUT, where the resource is fully replaced with a new one.
\item[TRACE:] echo the received request. This is useful to find out if a given HTTP request has been modified by intermediates (e.g., proxies) between the client and the server.
\item[OPTIONS:] list all available HTTP operations on the given resource. For example, it could be possible to GET a HTML file, but not DELETE it.
\item[CONNECT:] establish a tunneling connection through an HTTP proxy, usually needed for encrypted communications.
\end{description}

When a client sends an HTTP request, the server will send back an HTTP response with headers and possibly a payload in the body. 
Furthermore, the response will also contain a numeric, three digit status code.
There are five groups/families of codes, specified by the first digit:
\begin{description}[leftmargin=2em]
\item[1xx:] used for provisional responses, like confirming the switching of protocol (101) or that a previous, conditional request in which only the headers were sent should continue to send the body as well (100).
\item[2xx:] returned if the request was handled successfully (200). The server could for example further specify that a new resource was created (201), e.g., as a result of a POST command, or that nothing is expected in the response body (204), e.g., as a result of a DELETE command.
\item[3xx:] those codes are used for redirection, e.g., to tell the client that the requested resource is now available at a different location.
The redirection could be just temporary (307) or permanent (301).
\item[4xx:] used to specify that the user request was invalid (400). 
A typical case is requesting a resource that does not exist (404), or trying to access a protected resource without being authenticated (401) or authorized (403).
\item[5xx:] returned if the server cannot provide a valid response (500).
A typical case is if the code of the business logic has a bug, and an exception is thrown during the request processing, which is then caught by the application server (i.e., the whole server is not going to crash if an exception is thrown). 
However, this kind of code could also be returned if the needed external services (e.g., a database) are not responding correctly. For example, if the hardrive of a database breaks, a server could still be able to respond with a 500 code HTTP, even though it cannot use the database. 
\end{description}

The HTTP protocol is \emph{stateless}: each incoming request needs to provide all the information needed to be processed, as the HTTP protocol does not store any previous information. 
To maintain  state for a user doing several related HTTP requests (e.g., think about a shopping cart), then \emph{cookies} need to be employed: those are just HTTP headers with a unique id created by the server to recognize a given user. The user will need to include such header in all of his HTTP requests.

\subsection{REST}
For many years, the main way to write a web service was to use SOAP (Simple Object Access Protocol), which is a communication protocol using XML enveloped messages.
However, in recent years, there has been a clear shift in industry toward REST (Representational State Transfer) when developing web services.
All major players are now using REST, like for example
Google\footnote{https://developers.google.com/drive/v2/reference/},
Amazon\footnote{http://docs.aws.amazon.com/AmazonS3/latest/API/Welcome.html},
Twitter\footnote{https://dev.twitter.com/rest/public},
Reddit\footnote{https://www.reddit.com/dev/api/},
LinkedIn\footnote{https://developer.linkedin.com/docs/rest-api},
etc.

The concepts of REST were first introduced in a highly influential (nearly 6000 citations so far) PhD thesis~\cite{fielding2000architectural} in 2000.
REST is not a protocol (like SOAP is), but rather a set of architectural guidelines on building web services on top of HTTP.
Such client-server applications needs to satisfy some constraints to be considered RESTful, like being stateless and the resources should explicitly state if they are cacheable or not.  
Furthermore, resources should be identified with a URI. 
The representation of a resource (JSON or XML) sent to the client is independent from the actual format of the resource (e.g., a row in a relational database). 
These resources should be managed via the appropriate HTTP methods, e.g., a resource should be deleted with a DELETE request and not a POST one.

Let us consider an example of a RESTful web service that provides access to a product catalog. Possible available operations could be:
\begin{description}
\item[GET       /products] (return all available products)
\item[GET       /products?k=v] (return all available products filtered by some custom parameters)
\item[POST      /products] (create a new product)
\item[GET       /products/\{id\}] (return the product with the given id)
\item[GET       /products/\{id\}/price] (return the price of a specific product with a given id)
\item[DELETE    /products/\{id\}] (delete the product with the given id)
\end{description}

Note that those URIs do not specify the format of the representation returned. 
For example, a server could provide the same resource in different formats, like XML or JSON, and that should be specified in the headers of the request.

Another aspect of REST is the so called HATEOAS (Hypermedia As The Engine Of Application State),
where each resource representation should also provide links to other resources (in a similar way as links in web pages).
For example, when calling a \emph{GET /products}, not only all products should be returned, but also there should be links to what other methods are available.
Ideally, given the main entry point of an API, such API should be fully discoverable by using those links.
However, the use of HATEOS is quite rare in practice~\cite{rodriguez2016rest}, mainly due to the lack of a proper standard on how links should be defined (e.g., a JSON or XML schema), and the extra burden that it would entail on the clients.

Note: as REST is not a protocol, but just a set of architectural guidelines~\cite{fielding2000architectural}, the ubiquitous term ``REST'' has often been misused in practice. Many Web APIs over HTTP have been called and marketed as REST, although strictly speaking they cannot be considered as fully REST~\cite{rodriguez2016rest,fielding2000architectural}.   
Although in this paper we use the term REST, the presented novel technique would work as well to any Web API which is  accessed via HTTP endpoints, and where the payload data is expressed in a language like JSON or XML.

\subsection{Search-Based Software Testing}

Test data generation is a complex task, as software can be arbitrarily complex.
Furthermore, many developers find it tedious to write test cases.
Therefore, there has been a lot of research on how to automate the generation of high quality test cases. 
One of the easiest approach is to generate test cases at random~\cite{AIB11}.
Although it can be effective in some contexts, random testing is not a particularly effective testing strategy,
as it might cover just small parts of the tested software.
For example, it would not make much sense to use a naive random testing strategy on a RESTful API, as it would be extremely
unlikely that a random string would result in a valid, well-formed HTTP message.

Among the different techniques proposed throughout the years, search-based software engineering has been particularly effective at solving many different kinds of software engineering problems~\cite{harman2012search}, in particular software testing~\cite{ABHP09}, with advanced tools for unit test generation like 
EvoSuite\footnote{https://github.com/EvoSuite/evosuite}~\cite{fraser2011evosuite,fraser2014large}.
Software testing can be modeled as an optimization problem, where one wants to maximize the code coverage and fault detection of the generated test suites.
Then, once a fitness function is defined for a given testing problem, a search algorithm can be employed to explore the space of all possible solutions (test cases in this context).

There are several kinds of search algorithms, where Genetic Algorithms (GAs) are perhaps the most famous. 
In a GA, a population of individuals is evolved for several generations.
Individuals are selected for reproduction based on their fitness value, and then go through a crossover operator (mixing the material of both parents) and mutations (small changes) when sampling new offspring.
The evolution ends either when an optimal individual is evolved, or the search has run out of the allotted time.

\section{Related Work}

Canfora and Di Penta provided a discussion on the trends and challenges of SOA testing~\cite{canfora2006testing}. 
Afterwards, they provided a more in detail survey~\cite{canfora2009service}.
There are different kinds of testing for SOA (unit, integration, regression, robustness, etc.), which also depend on which stakeholders are involved, e.g., service developers, service providers, service integrators and third-party certifiers.
Also Bertolino et al.~\cite{bertolino2012trends} 
discussed the trends and challenges in SOA validation and verification.

Successively, Bozkurt et al.~\cite{bozkurt2013testing} carried out a survey as well on SOA testing, in which 177 papers were analysed. 
One of the interesting results of this survey is that, although the number of papers on SOA testing has been increasing throughout the years, only 11\% of those papers provide any empirical study on actual, real systems. In 71\% of the cases, no experimental result at all was provided, not even on toy case studies.

A lot of the work in the literature has been focusing on black-box testing of SOAP web services described with WSDL (Web Services Description Language).
Different strategies have been proposed, like for example~\cite{
xu2005testing,
bai2005wsdl,
martin2006automated,
ma2008wsdl,bartolini2009ws,li2016generating}.
If those services also provide a semantic model (e.g., in OWL-S format), that can be exploited to create more ``realistic'' test data~\cite{bozkurt2011automatically}.
When in SOAs the service compositions are described with BPEL (Web Services Business Process Execution Language), different techniques can be used to generate tests for those compositions~\cite{wotawa2013fifty,jehan2014soa}

Black-box testing has its advantages, but also its limitations. Coverage measures could improve the generation of tests but, often, web services are remote and there is no access to their source code. 
For testing purposes, Bartolini et al.~\cite{bartolini2011whitening} proposed an approach in which feedback on code coverage is provided as a service, without exposing the internal details of the tested web services.
However, the insertion of the code coverage probes had to be done manually.
A similar approach has been developed by Ye and Jacobsen~\cite{ye2013whitening}.
In our approach in this paper, we do provide as well code coverage as a service, but our approach is fully automated (e.g., based on on-the-fly bytecode manipulation).

Regarding RESTful web services, Chakrabarti and Kumar~\cite{chakrabarti2009test} provided a testing framework
in which ``automatic generation test cases corresponding to an exhaustive list of all valid combinations of query parameter values''. 
Seijas et al.~\cite{lamela2013towards} proposed a technique to generate tests for RESTful API based on an idealised, property-based  test model.
Chakrabarti and Rodriquez~\cite{chakrabarti2010connectedness} defined a technique to formalize the ``connectedness'' of a RESTful service, and generate tests based on such model.
When formal models are available, techniques like in~\cite{pinheiro2013model} and in~\cite{fertig2015model} can be used as well.
Our technique is significantly different from those approaches, as it does not need the presence of any formal model, can automatically exploit white-box information, and uses an evolutionary algorithm to guide the generation of effective tests.

Regarding the usage of evolutionary techniques for testing web services, Di Penta et al.~\cite{di2007search} proposed an approach for testing Service Level Agreements (SLA).
Given an API in which a contract is formally defined stating ``for example, that the service provider guarantees to the service consumer a response time less than 30 ms and a resolution greater or equal to 300 dpi'', an evolutionary algorithm is used to generate tests to break those SLAs. The fitness function is based on how far a test is from breaking the tested SLA, which can be measured after its execution.

\section{Proposed Approach}

In this paper, we propose a novel approach to automatically generate test cases for RESTful API web services.
We consider the testing of a RESTful service in isolation, and not as part of an orchestration of two or more services working together (e.g., like in a microservice architecture).
We consider the case in which our approach to automatically generate test cases is used directly by the developers of such RESTful services.
As such, we assume the availability of the source code of these developed services.
The goal is to generate test cases with high code coverage and that can detect faults in the current implementation of those services.
We hence need to define how a test case looks like, what can be used as an automated oracle (needed to check for faulty behaviours), and how a search algorithm can be used to generate such tests.

\subsection{Test Case}

In our context, a test case is one or more HTTP requests towards a RESTful service.
The test data can hence be seen as a string, representing the HTTP request.
However, besides the given structure of a HTTP request (e.g., headers and parameters in the resource paths), such test data can be arbitrarily complex. 
For example, the content in the body section could be in any format.
As currently JSON is the main format of communication in RESTful APIs, in this paper we will focus just on such format.
Handling other less popular formats, like for example XML, would be just a matter of engineering effort.

At any rate, before being able to make an HTTP request, we need to know what API methods are available.
In contrast to SOAP, which is a well defined protocol, REST does not have a standard to define the available APIs.
However, a very popular tool for REST documentation is 
Swagger\footnote{http://swagger.io},
which is currently available for more than 25 different programming languages.
Another tool is 
RAML\footnote{http://raml.org},
but it is less popular.
When a RESTful API is configured with Swagger, it will automatically provide a JSON file as a resource that will fully define which APIs are available in that RESTful service.
Therefore, the first step, when testing such RESTful service, is to retrieve such Swagger JSON definition.

\begin{figure}[!t]
\caption{\label{fig:swagger_get_put}
Swagger JSON definition of two operations (GET and PUT) on 
the \emph{/v1/activities/\{id\}} resource.}
\begin{lstlisting}[]
"/v1/activities/{id}": {
      "get": {
        "tags": [
          "activities"
        ],
        "summary": "Read a specific activity",
        "description": "",
        "operationId": "get",
        "produces": [
          "application/json"
        ],
        "parameters": [
          {
            "name": "id",
            "in": "path",
            "required": true,
            "type": "integer",
            "format": "int64"
          },
          {
            "name": "attrs",
            "in": "query",
            "description": "The attributes to include in the response. Comma-separated list.",
            "required": false,
            "type": "string"
          }
        ],
        "responses": {
          "default": {
            "description": "successful operation"
          }
        }
      },
      "put": {
        "tags": [
          "activities"
        ],
        "summary": "Update an activity with new information. Activity properties not specified in the request will be cleared.",
        "description": "",
        "operationId": "update",
        "produces": [
          "application/json"
        ],
        "parameters": [
          {
            "name": "id",
            "in": "path",
            "required": true,
            "type": "integer",
            "format": "int64"
          },
          {
            "in": "body",
            "name": "body",
            "required": false,
            "schema": {
              "$ref": "#/definitions/ActivityProperties"
            }
          }
        ],
        "responses": {
          "200": {
            "description": "successful operation",
            "schema": {
              "$ref": "#/definitions/Activity"
            }
          }
        }
      }
\end{lstlisting}
\end{figure}

Figure~\ref{fig:swagger_get_put} shows an extract from a Swagger definition of one of the systems we will use in the empirical study.
The full JSON file is more than 2000 lines of code.
In that figure, there is the definition for two HTTP operations (GET and PUT) on the same resource.
To execute a GET operation on such resource, there is the need of two values:
a numeric ``id'' which will be part of the resource path, and an optional query parameter
called ``attrs''.
For example, given the template 
\emph{/v1/activities/\{id\}}, 
one could make a request for 
\emph{/v1/activities/5?attrs=x}.

\begin{figure}[!t]
\caption{\label{fig:swagger_object}
Swagger JSON definition of a complex object type.}
\begin{lstlisting}[]
"ActivityProperties": {
  "type": "object",
  "properties": {
    "id": {
      "type": "integer", "format": "int64"}, 
    "name": {
      "type": "string", 
      "minLength": 0, "maxLength": 100 }, 
    "date_published": {
      "type": "string", "format": "date-time"},
    "date_created": {
      "type": "string", "format": "date-time"},
    "date_updated": {
      "type": "string", "format": "date-time"},
    "description_material": {
      "type": "string",
      "minLength": 0, "maxLength": 20000},
    "description_introduction": {
      "type": "string",
      "minLength": 0, "maxLength": 20000},
    "description_prepare": {
      "type": "string",
      "minLength": 0, "maxLength": 20000},
    "description_main": {
      "type": "string",
      "minLength": 0, "maxLength": 20000},
    "description_safety": {
      "type": "string",
      "minLength": 0, "maxLength": 20000},
    "description_notes": {
      "type": "string",
      "minLength": 0, "maxLength": 20000},
    "age_min": {
      "type": "integer", "format": "int32",
      "maximum": 100.0},
    "age_max": {
      "type": "integer", "format": "int32",
      "maximum": 100.0},
    "participants_min": {
      "type": "integer", "format": "int32"},
    "participants_max": {
      "type": "integer", "format": "int32"},
    "time_min": {
      "type": "integer", "format": "int32"},
    "time_max": {
      "type": "integer", "format": "int32"},
    "featured": {
      "type": "boolean", "default": false},
    "source": {
      "type": "string"},
    "tags": {
      "type": "array",
      "xml": {"name": "tag", "wrapped": true},
      "items": {"$ref": "#/definitions/Tag"}},
    "media_files": {
      "type": "array",
      "xml": {"name": "mediaFile", "wrapped": true},
      "items": {"$ref": "#/definitions/MediaFile"}},
    "author": {"$ref": "#/definitions/User"},
    "activity": {"$ref": "#/definitions/Activity"}
  }
}
\end{lstlisting}
\end{figure}

The PUT operation needs as well an ``id'' value, but not the optional parameter ``attrs''.
However, in its HTTP body, it can have a JSON representation of the resource to replace,
which is called ``ActivityProperties'' in this case.
Figure~\ref{fig:swagger_object} shows such object definition.
This object has many fields of different types, like numeric (e.g., ``id''), 
strings (e.g., ``name''), dates (e.g., ``date\_published''), arrays (e.g., ``tags'')
and other objects as well (e.g., ``author'').
When a test case is written for such PUT operation, besides specifying an ``id'' in the path, one would also need to instantiate such ``ActivityProperties'' object, and marshall it as a JSON string to add in the body of the HTTP request.

\subsection{Oracle}

\begin{figure}[!t]
\caption{\label{fig:rest_bug}
An example (in Java, using DropWizard) of endpoint definition to handle a GET request, where requesting a missing resource, instead of resulting in a 404 code, does lead to a 500 code due to a null pointer exception.}
\begin{lstlisting}[language=java]
@GET @Timed
@Path("{id}/file")
@Produces(MediaType.APPLICATION_OCTET_STREAM)
@UnitOfWork
@ApiOperation(value = "Download media file. Can resize " +
  "images (but images will never be enlarged).")
public Response downloadFile(
    @PathParam("id") long id,
    @ApiParam(value = "" +
      "The maximum width/height of returned images. " +
      "The specified value will be rounded up to the " +
      "next 'power of 2', e.g. 256, 512, 1024 and so on.")
    @QueryParam("size") int size) 
    {
    MediaFile mediaFile = dao.read(id);
    try {
        URI sourceURI = new URI(mediaFile.getUri());
        ...
    } catch (IOException e) {
        ...
    }
    ...
\end{lstlisting}
\end{figure}

When automatically generating test cases with a white-box approach, like for example trying to maximize statement coverage, there is the problem of what to use as an automated 
\emph{oracle}~\cite{barr2015oracle}.
An oracle can  be consider as a function that tells whether the result of a test case is correct or not.
In manual testing, the developers decide what should be the expected result for a given test case, and write such expectation as an assertion check directly in the test cases.
In automated test generation, where many hundreds if not thousands of test cases are generated, asking the developers to write such assertions is not really a viable option.

There is no simple solution for the oracle problem, just different approaches with different degrees of success and limitations~\cite{barr2015oracle}.
In system-level testing, the most obvious automated oracle is to check if the whole system under test does crash (e.g., a segmentation fault in C programs) when a test is executed.
Such test case would have detected a bug in the software, but not all bugs lead to a complete crash of an application (likely, just a small minority of bugs are of this kind).
Another approach is to use formal specifications (e.g., pre/post conditions) as automated oracles, but those are seldom used in practice.

In unit testing, one can look at thrown exceptions in the tested classes/methods~\cite{evo1600faults2015}.
However, one major problem here is that, often, thrown exceptions are not a symptom of a bug, but rather a violation of an unspecified pre-condition (e.g., inserting a null input when the target function is not supposed to work on null inputs).

Even if no automated oracle is available, generated tests are still useful for \emph{regression testing}. 
For example, if a tested function $foo$ takes as input an integer value, and then returns an integer as result of the computation, then an automatically generated test could capture the current behavior of the function in an assertion, for example:
\begin{lstlisting}[language=java]
int x = 5;
int res = foo(x);
assertEquals(9, res);
\end{lstlisting}
Now, a test generation tool could choose to create a test with input value $x=5$, but it would not be able to tell if the expected output should really be $9$ (which is the actual value returned when calling $foo(x)$).
A developer could look at such generated test, and then confirm if indeed $9$ is the expected output.
But, even if he doesn't check it, such test could be added to the current set of test cases, and then run at each new code change as part of a Continuous Integration process (e.g., Jenkins\footnote{https://jenkins.io}).
If a modification to the source code leads $foo$ to return a different values than $9$ when  called with input $5$, then that test case will fail.
At this point, the developers would need to check if indeed the recently introduced change does break the function (i.e., it is a bug), or rather if the semantics of that function has changed. 

In the case of test cases for RESTful APIs, the generated tests can be used for regression testing as well.
This is also particularly useful for security: for example, a HTTP invocation in which the returned status is 403 (unauthorized) can detect regression faults in which the authorization check are wrongly relaxed.
Furthermore, the status codes can be used as automated oracles.
A 4xx code does not mean a bug in the RESTful web service, but a 5xx can.
If the environment (e.g., databases) of the tested web service is working correctly, then a 5xx status code would often mean a bug in such service.
A typical example is thrown exceptions: the application server will not crash if an exception is thrown in the business logic of a RESTful endpoint.
Such exception would be caught, and a 5xx code (e.g., 500) would be returned.
Note: if the user sends invalid inputs, he should get back a 4xx code, not a 5xx one.
Not doing input validation and letting the endpoint throwing an exception would have two main problems:
\begin{itemize}
\item the user would not know that it is his fault, and so just think it is a bug in the web service. Inside an organisation, such developer might end up wasting time in filling a bug report, for example. Furthermore, the 5xx code would not give him any hint on how to fix how he is calling the RESTful API. 
\item the RESTful endpoint might do a sequence of operations on external resources (e.g., databases and other web services) that might require to be atomic. If an exception is thrown due to a bug after some, but not all, of those operations are completed, the environment might be left in a inconsistent state, making the entire system working incorrectly.
\end{itemize}

Figure~\ref{fig:rest_bug} shows a simple example of endpoint definition which contains bugs.
This code is from one of the projects used in the empirical study.
In that particular case, a resource (a media file) is referenced by id in the endpoint path (i.e., \emph{@Path(``\{id\}/file'')}).
Such id is used to load such resource from a database (i.e., \emph{dao.read(id)}), but there is no check if it exists (e.g., if different from \emph{null}).
Therefore, when a test is created with an invalid id, the statement \emph{mediaFile.getUri()}
does result in a null pointer exception.
Such exception is propagated to the application server (Jetty, in this case), which will create a HTTP response with status 500.
The expected, correct result here should had been a 404 (not found) code.

\subsection{Code Instrumentation}

To generate high coverage test cases, coverage itself needs to be measured.
Otherwise, it would not be possible to check if a test has higher coverage than another one.
So, when the system under test (SUT) is started, it needs to be \emph{instrumented} to collect code coverage metrics.
How to do it will depend on the programming language.
In this paper, for our prototype, we started by focusing on Java.

Coverage metrics can be collected by automatically adding probes in the SUT.
This is achieved by instantiating a Java Agent that intercepts all class loadings,
and then add probes directly in the bytecode of the SUT classes.
This process can be fully automated by using libraries like 
\emph{ea-agent-loader}\footnote{https://github.com/electronicarts/ea-agent-loader}
(for Java Agent handling) and
\emph{ASM}\footnote{http://asm.ow2.org/}
(for bytecode manipulation).
Such an approach is the same used in unit test generation tools for Java like
EvoSuite~\cite{fraser2011evosuite}.

Measuring coverage is not enough.
Knowing that a test case cover 10\% of the code does not tell us how more code could be covered.
Often, code is not covered because it is inside blocks guarded by \emph{if} statements with complex predicates.
Random input data is unlikely to be able to solve the constraints in such complex predicates.
This is a very well known problem in search-based unit testing~\cite{Mcm04}. 
A solution to address this problem is to define \emph{heuristics} that measure how far a test data is to solve a constraint.
For example, given the constraint $x==0$, although neither $5$ nor $1000$ does solve such constraint, the value $5$ is \emph{heuristically} closer than $1000$ to solve it.
The most famous heuristic in the literature is the so called \emph{branch distance}~\cite{Kor90,Mcm04}. 
In our approach, we use the same kind of branch distance used in unit testing, by automatically instrumenting the boolean predicates when the bytecode of a class is loaded for the first time (same way as for the code coverage probes).

Even if one can measure code coverage and branch distances by using bytecode manipulation (e.g., for JVM languages), there is still the question of how to retrieve such values.
In unit testing, a test data generation tool would run in the same process of where the tests are evaluated, and so such values could be directly read.
It would be possible to do the same for system testing:
the testing tool and the SUT could run in the same process, e.g., the same JVM.
However, such an approach is not optimal, as it would limit the applicability of a test data generation tool to only RESTful services written in the same language.
Furthermore, there could be third-party library version conflicts between the testing tool and the SUT.
As the test cases would be independent of the language in which a RESTful API is written (as they are just HTTP calls),
focusing on a single language is an unnecessary limitation.

Our solution is to have the testing tool and the SUT running in different processes.
For when the SUT is run, we provide a library with functionalities to automatically instrument the SUT code.
Furthermore, the library itself would automatically provide a RESTful API to export all the coverage and branch distance information in a JSON format.
The testing tool, when generating and running tests, would use such API to determine the fitness of these tests.
The testing tool would be just one, but, then, for each target programming language (e.g., Java, C\# and JavaScript) we would just need its library implementation for the code instrumentation.

Such an approach would not work well with unit testing: the overhead of an HTTP call to an external process would be simply too great compared to the cost of running a unit test.
On the other hand, in system-level testing, an entire application (a RESTful web service in our case) runs at each test execution.
Although non-zero, such overhead would be more manageable, especially when the SUT itself has complex logic and interacts with external services (e.g., a database).

Although the overhead of instrumentation is more manageable, it still needs to be kept under control.
In particular, in our approach we consider the two following optimizations:
\begin{itemize}
\item when the SUT starts, the developer has to specify which packages to instrument.
Instrumenting all the classes loaded when the SUT starts would be far too inefficient.
For example, there is no point in collecting code coverage metrics on third-party libraries, like the application servers (e.g., Jetty or Tomcat), or ORM libraries like Hibernate.
\item by default, when querying the SUT for code coverage and branch distance information, not all information is retrieved: only the one of newly covered targets, or better branch distance, is returned. 
The reason is that, if the SUT is 100 thousand lines of code, then you do not want to un/marshal JSON data with 100 thousand elements at each single test execution.
The testing tool will ask explicitly for which testing targets it needs information for.
For example, if a target is fully covered with an existing test, there is no point in collecting info for that target when generating new tests aimed at covering the other remaining targets. 
\end{itemize}

\subsection{Search Algorithm}

Given a way to define a test case, run it, and collect metrics on its performance (e.g., code coverage and branch distances), then we could use any search algorithm.
In our approach, we evaluate the use of a Genetic Algorithm (GA) to generate test cases.

The final output of our technique is a test suite, which is a collection of test cases.
Each test case will cover one or more testing targets.
In our case, we consider two types of testing targets:
(1) coverage of statements in the SUT; 
and (2) returned HTTP status codes for the different API endpoints (i.e., we want to cover not only the happy day scenarios like 2xx, but also user errors and server errors, regardless of the achieved coverage).

As we need to evolve test suites, we use the \emph{Whole Test Suite} approach~\cite{GoA_TSE12}, with the extra usage of a test archive~\cite{rojas2016detailed}.
A GA individual will be a set of test cases, randomly initialized, with variable size and length.
The fitness of a test suite is the aggregated fitness of all of its test cases.
The crossover operator will mix test cases from two parent sets when new offspring are generated.
The mutation operator will do small modifications on each test case, like increasing or decreasing  a numeric variable by 1.

We support all valid types in JSON (e.g., numbers, strings, dates, arrays and objects).
Some of them are treated specially.
For example, for date times, as genotype we consider  an array of six bounded numeric values:
year, month, day, hour, minute and seconds.
We consider valid values (e.g., seconds are from 0 to 59), but also some invalid ones (e.g., -1 second) to check how the SUT behaves when handling time stamps with invalid format.
When such date is used in a JSON variable, the phenotype will be a date string composed from those six integer values.

When a test is executed, we check all targets it covers. If it covers a new target, the test will be copied from the test suite and added to an archive, to  not lose it during the search (e.g., due to a mutation operation in the next generations).
At the end of the search, we collect all tests stored in the archive, remove the redundant ones, and write the minimised suite to disk as a test class file.

\subsection{Tool Implementation}

We have implemented a tool prototype in Kotlin to experiment with the novel approach discussed in this paper.
The tool is called \evo, and it is released under the LGPL open-source license.

For handling the SUT (start/stop it, and code instrumentation), we have developed a library for Java, which in theory should work for any JVM language (Java, Kotlin, Groovy, Scala, etc.).
However, we have tried it only on Java systems.
Our tool prototype can output test cases in different formats, like JUnit 4 and 5, in both Java and Kotlin.
Such test suites will be fully self-contained, i.e., they will also deal with the starting/stopping of the SUT.
The test cases are configured in a way that the SUT is started on an ephemeral TCP port, which is an essential requirement for when tests are run in parallel (i.e., to avoid a SUT trying to open a TCP port that is already bound).
The generated tests can be directly called from an IDE (e.g., IntelliJ or Eclipse), and can be added as part of a Maven or Gradle build.

In the generated tests, to make the HTTP calls toward the SUT, we use the highly popular 
RestAssured\footnote{https://github.com/rest-assured/rest-assured} 
library.
Assertions are currently generated only for the returned HTTP status codes.

\subsection{Manual Preparations}
\label{sub:manual}

\begin{figure}[!t]
\caption{\label{fig:rest_controller}
Example of class that needs to be implemented by the developers of the SUT to enable the usage of our test case generation tool.
In this particular case, the SUT is written with Spring, where \emph{Application} is the main entry point of the SUT.
}
\begin{lstlisting}[language=java]
public class EMController extends RestController {

    private ConfigurableApplicationContext ctx;
    private final int port;
    private Connection connection;

    public EMController(){this(0);}
    public EMController(int port) {
        this.port = port;
    }

    @Override public int getControllerPort(){
        return port;
    }

    @Override public String startSut() {

        ctx = SpringApplication.run(Application.class,
                new String[]{"--server.port=0"});

        if(connection != null){
            try { connection.close();
            } catch (SQLException e) {
                e.printStackTrace();
            }
        }
        JdbcTemplate jdbc = ctx.getBean(
                                   JdbcTemplate.class);
        try {
            connection = jdbc.getDataSource()
                             .getConnection();
        } catch (SQLException e) {
            e.printStackTrace();
        }
        return "http://localhost:"+getSutPort();
    }

    protected int getSutPort(){
        return (Integer)((Map) ctx.getEnvironment()
                .getPropertySources().get("server.ports")
                .getSource()).get("local.server.port");
    }

    @Override public boolean isSutRunning() {
        return ctx!=null && ctx.isRunning();
    }

    @Override public void stopSut() { ctx.stop();}

    @Override public String getPackagePrefixesToCover() {
        return "org.javiermf.features.";
    }

    @Override public void resetStateOfSUT() {
        ScriptUtils.executeSqlScript(connection, 
            new ClassPathResource("/empty-db.sql"));
        ScriptUtils.executeSqlScript(connection, 
            new ClassPathResource("/data-test.sql"));
    }

    @Override public String getUrlOfSwaggerJSON() {
        return "http://localhost:"+getSutPort()+
           "/swagger.json";
    }

    @Override public List<AuthenticationDto> 
                        getInfoForAuthentication(){
        return null;
    }
}
\end{lstlisting}
\end{figure}

In contrast to tools for unit testing like EvoSuite, which are 100\% fully automated (a user just need to select for which classes tests should be generated),
our tool prototype for system/integration testing of RESTful APIs does require some manual configuration.

The developers of the RESTful APIs need to import our library, and then create a class that extends the \emph{RestController} class in such library.
The developers will be responsible to define how the SUT should be started, where the Swagger schema can be found, which packages should be instrumented, etc.
This will of course vary based on how the RESTful API is implemented, e.g.,
if with 
Spring\footnote{https://github.com/spring-projects/spring-framework}, 
DropWizard\footnote{https://github.com/dropwizard/dropwizard}, 
Play\footnote{https://github.com/playframework/playframework}, 
Spark\footnote{https://github.com/perwendel/spark} 
or JEE.

Figure~\ref{fig:rest_controller} shows an example of one such class we had to write for one of the SUTs in our empirical study.
That SUT uses Spring. 
That class is quite small, and needs to be written only once.
It does not need to be updated when there are changes internally in the API.
The code in the superclass \emph{RestController} will be responsible to do the automatic bytecode instrumentation of the SUT, and it will also start a RESTful service to enable our testing tool to remotely call the methods of such class.

However, besides starting/stopping the SUT and providing other information (e.g., location of the Swagger file), there are two further tasks the developers need to perform:
\begin{itemize}
\item	RESTful APIs are supposed to be stateless (so they can easily scale horizontally), but they can have side effects on external actors, such as a database.
In such cases, before each test execution, we need to reset the state of the SUT environment.
This needs to be implemented inside the \emph{resetStateOfSUT()} method.
In the particular case of the class in Figure~\ref{fig:rest_controller},
two SQL scripts are executed: one to empty the database, and one to fill it with some existing values.
We did not need to write those scripts by ourself, as we simply re-used the ones already available in the manually written tests in that SUT.
How to automatically generate such scripts would be an important topic for future investigations.
\item if a RESTful API requires some sort of authentication and authorization, such information has to be provided by the developers in the \emph{getInfoForAuthentication()} method.
For example, even if a testing tool would have full access to the database storing the passwords for each user, it would not be possible to reverse engineer those passwords from the stored hash values. 
Given a set of valid credentials, the testing tool will use them as any other variable in the test cases, e.g., to do HTTP calls with and without authentication.
\end{itemize}

\section{Empirical Study}

In this paper, we have carried out an empirical study aimed at answering the following research questions.

\begin{description}
\item[{\bf RQ1}:] Can our technique automatically find real faults in existing RESTful web services?
\item[{\bf RQ2}:] How do our automatically generated tests compare, in terms of code coverage, with the already existing, manually written tests?
\item[{\bf RQ3}:] What are the main factors that impede the achievements of better results?
\end{description}

\subsection{Artefact Selection}

To achieve sound, reliable conclusions from an empirical study,
ideally we would need a large set of artefacts for experimentation, selected in an unbiased way~\cite{FrA12b}.
However, for this paper, this was not really possible.
First, system level testing requires some manual configuration (e.g., recall Section~\ref{sub:manual}).
Second, our novel prototype is still in a early stage of development, and it might not be ready yet to handle many different types of systems.
Third, a major issue is that RESTful web services, although extremely popular among enterprises in industry, are less common among open-source projects.
Finding the right projects that do not require complex installations (e.g., special databases and connections to third-party tools) to run is not a trivial task.

\begin{table*}[!t]
\centering
\caption{\label{tab:case_study} Information about the three RESTful web services used in the empirical study. We report their number of Java classes and lines of code, also for their tests. 
We also specify the number of endpoints, i.e., the number of exposed resources and HTTP methods applicable on them, as well as if they do access a database or external web services.
}
\begin{tabular}{l rr rr r cc} 
\toprule
Name & \# Classes & LOCs & \# Test Classes & Test LOCs & Endpoints & Database & Ext. Services \\			
\midrule
\emph{FeaturesService} & 23 & 1247 & 14 &  822 & 18 & Yes & No \\
\emph{Industrial}      & 50 & 3584 & 13 & 2313 & 10 & Yes & Yes\\
\emph{ScoutApi}        & 75 & 7479 & 21 & 2428 & 49 & Yes & No \\
\bottomrule
\end{tabular}
\end{table*}

We used 
Google BigQuery\footnote{https://cloud.google.com/bigquery}
to analyse the content of the Java projects hosted on
GitHub\footnote{https://github.com},
which is the main repository for open-source projects.
We searched for Java projects using Swagger.
We excluded too large projects (as potentially too difficult to handle at this early stage), as well as the too small, trivial ones.
We dowloaded and tried to compile and run several of these projects, with different degrees of success.
In the end, for the empirical study in this paper, we manually chose two different RESTful web services which we could compile and run their test cases with no problems. 
These services are called 
\emph{FeaturesService}\footnote{https://github.com/JavierMF/features-service}
and \emph{ScoutApi}\footnote{https://github.com/mikaelsvensson/scout-api}.
Besides those two open-source web services, we also used a RESTful web service provided by one of our industrial partners.
Due to non-disclosure agreements, we can only provide limited information about that particular web service. 
Data about these three RESTful web services is summarized in Table~\ref{tab:case_study}.

Those three RESTful web services contain between 2 and 10 thousand lines of codes (tests included).
This is a typical size, especially in a microservice architecture~\cite{newman2015building}.
The reason is that, to avoid the issues of monolithic applications, such services usually become split if growing too large, as to make them manageable by a single, small team.
This, however, does also imply that enterprise applications can end up being composed of hundreds of different services. 
In this paper, we focus on the testing of RESTful web services in isolation, and not their orchestration in a whole enterprise system.

\subsection{Experiment Settings}

On each of the web services in our case study, we ran our tool to generate test cases.
As our technique is based on randomized algorithms, each experiment has been repeated 30 times with different random seeds.

The longer a search algorithm is run, the better results one can expect.
For the experiments in this paper, we use a stopping criterion of 100 thousand fitness evaluations.
On the computer used to run these experiments, each run took roughly between two and four minutes.
Considering three web services, and 30 repetitions per service, we ran a total of 90 experiments, for a total of 9 million fitness evaluations.

In a search algorithm, there can be many parameters that need to be configured. 
For example, in a GA, one has to specify the population size, the probability of applying the crossover operator, the probability of applying mutation, etc.
The choice of those parameters will impact the performance of the search algorithm.
In \emph{parameter tuning}, one would experiment with different parameter settings to find the best ones for the given domain (e.g., test case generation for RESTful API).
Fortunately, not only there are general guidelines on how to choose those settings, but search algorithms are often robust: using non-tuned, default settings from the literature can still provide good enough results on average~\cite{arcuri2013parameter}.  

For the GA employed in this paper, we used a population size of 30, crossover probability $0.7$, and mutation probability $1/n$ per variable in a test (where $n$ is the number of variables, so on average only one will be mutated). When a test suite is generated, it will have a random number of tests between 1 and 30.

\subsection{Experiment Results}
\label{sec:results}

\begin{table}[!t]
\centering
\caption{\label{tab:results}
Results of the experiments, based on 30 runs per SUT. In particular, we report the average number of test cases in the final test suites, the average number of distinct endpoints with at least one test leading to a 5xx status code response, the average number of different HTTP status responses per endpoint, and their max number for a single endpoint.}
\begin{tabular}{ l r rrr } 
\toprule 
SUT & \#Tests & \#5xx & \#Codes & Max \\ 
\midrule 
\emph{FeaturesService} & 46.0 & 15.0 & 2.6 & 4\\ 
\emph{Industrial} & 25.4 & 6.0 & 2.5 & 3\\ 
\emph{ScoutAPI} & 105.3 & 18.0 & 2.0 & 3\\ 
\bottomrule 
\end{tabular} 

\end{table}

\begin{figure}[!t]
\caption{\label{fig:test}
Generated RestAssured test (Java, JUnit 4) for the endpoint shown in Figure~\ref{fig:rest_bug}.
We also show the scaffolding code used to automatically start/stop/reset the SUT.
}
\begin{lstlisting}[language=java]
static EMController controller = new EMController();
static String baseUrlOfSut;

@BeforeClass
public static void initClass() {
    baseUrlOfSut = controller.startSut();
    assertNotNull(baseUrlOfSut);
}

@AfterClass
public static void tearDown() {
    controller.stopSut();
}

@Before
public void initTest() {
    controller.resetStateOfSUT();
}

@Test
public void test0() throws Exception {
    
    given().header("Authorization", "ApiKey user")
            .accept("*/*")
            .get(baseUrlOfSut + 
               "/api/v1/media_files/-4203492812/file" + 
               "?size=-141220")
            .then()
            .statusCode(500);
    }
}
\end{lstlisting}
\end{figure}

Table~\ref{tab:results} shows the results of the experiments on the three different RESTful web services.
Although during the search we evaluated 100 thousand HTTP calls per run, the final test suites are much smaller, on average between 25 (\emph{Industrial}) and 105 (\emph{ScoutAPI}) tests. 
This is because we only keep tests that contribute to cover our defined testing targets (i.e., code statements and HTTP return statuses per endpoint).

These tests, on average, can lead the SUTs to return 5xx status code responses in 39 distinct cases.
In the case of \emph{FeaturesService} (15) and \emph{ScoutAPI} (18), those tests pointed  to actual bugs in those systems.
A simple example is the one we previously showed in Figure~\ref{fig:rest_bug}.
A generated test revealing such bug is shown in Figure~\ref{fig:test}.
Note: for that particular endpoint, there are only three decisions to make: (1) whether or not to call it with a valid authentication header; (2) the numeric value of the ``id'' in the resource path; and (3) the numeric value of the ``size'' query parameter.

However, not all the test cases resulting in a 5xx response in the \emph{Industrial} web service were revealing bugs.
In one case, a 500 returned code was the expected, correct behavior.
This happens because the \emph{Industrial} web service, in contrast to the other two, does access some external web services.
Usually, in such a testing context, one would use a tool like 
WireMock\footnote{http://wiremock.org}
to mock out the responses of such external web services.
However, this was not configured for our experiments.
Therefore, every time the SUT tried to connect to such external web services, those calls failed, and the SUT could not complete its operations.
This is the type of situations where a 500 code is the right response, although there is no bug in the SUT.

In Table~\ref{tab:results} we can see that, on average, each single endpoint is called with data that result in at least two different status codes.
In some cases, it even happened that the same endpoint was called from test data that resulted in four different returned status codes. 

\begin{result} 
{\bf RQ1:} Our novel technique automatically found 38 real bugs in the analyzed web services.
\end{result}

\begin{table}[!]
\centering
\caption{\label{tab:coverage}
Statement coverage results for the generated tests compared to the ones obtained by the already existing, manually written tests.
}
\begin{tabular}{l rr}
\toprule
SUT & Coverage & Manual Cov. \\
\midrule
\emph{FeaturesService} & 41\% & 82\% \\
\emph{Industrial}      & 18\% & 47\% \\
\emph{ScoutAPI}        & 20\% & 43\% \\
\bottomrule
\end{tabular}
\end{table}

\phantom{a}

Besides finding faults, the generated test suites can also be used for regression testing.
To be useful in such context, it would desirable that such test suites would have high code coverage.
Otherwise, a regression in a non-executed statement would not fail any of the tests.
Table~\ref{tab:coverage} shows the statement coverage results of the generated tests.
Such results are compared against the ones of the already existing, manually written tests.
Code coverage was measured by running the tests directly from the IDE IntelliJ, using its code coverage tools.
This also helped to check if the generated tests worked properly. 
The results in Table~\ref{tab:coverage} clearly show that, for the generated tests, the obtained code coverage is  lower.

\begin{result} 
{\bf RQ2:} On average, the generated test suites obtained between 18\% and 41\% statement coverage. This is lower than the coverage of the existing test cases in those SUTs.
\end{result}

\subsection{Discussion}

The results in Section~\ref{sec:results} clearly show that our novel technique is useful for software engineers, as it can automatically detect real faults in real systems.
However, albeit promising, code coverage results could had been better.
Therefore, we did manually analyze some of the cases in which only low coverage was obtained. 
We found out at least three main reasons for those results, and so here we discuss possible solutions to improve performance even further:
\begin{description}
\item[String Constraints:]
Some branches in the SUTs depend on string constraints. 
Strings are complex to handle, and, throughout the years, for unit testing different techniques based on specialized search operators~\cite{Alshraideh06} and seeding strategies~\cite{rojas2016seeding} have been proposed.
Our prototype does not support such techniques yet.
It will be a matter of implementing and adapting them, and, then, evaluate if they do perform well in our testing context.
\item[Databases:] even if a database is initialised with valid data, our technique has currently no way to check what is inside them, even less, it cannot generate or modify such data. Recall the example of Figure~\ref{fig:rest_bug}: even if there is valid data in the database, our testing tool has no gradient toward generating an id matching an existing key in the database.
The testing tool should be extended to be able to check all SQL commands executed by the SUT, and use such info when generating the HTTP calls. 
Furthermore, the generation of data in the database should be part of the search as well:
a test case would not be any more just HTTP calls, but also SQL commands. 
\item[External Services:] like for databases, we need to handle accesses to external web services as well. This means using tools like WireMock, which should then be configured in the generated tests, and become part of the search.
\end{description}

\begin{result} 
{\bf RQ3:} String constraints, accesses to databases and external web services are the current main impediments. 
\end{result}

\section{Threats to Validity}

Threats to internal validity come from the fact that our empirical study is based on a tool prototype.
Faults in such tool might compromise the validity of our conclusions.
Although such prototype has been carefully tested, we cannot provide any guarantee that it is bug-free.
Furthermore, as our techniques are based on randomized algorithms, such randomness might affect the results.
To mitigate such problem, each experiment was repeated 30 times with different random seeds.

Threats to external validity come from the fact that only three RESTful web services were used in the empirical study.
Although those three services are not trivial (i.e., between 2 and 10 thousand lines of code), we cannot generalize our results to other web services.
However, besides open-source projects, we used an industrial one as well, which helps us increase our confidence that our novel technique can be helpful for practitioners.

\section{Conclusion}

RESTful web services are popular in industry.
Their ease of development, deployment and scalability make them one of the key tools in modern enterprise applications.
This is particularly the case when enterprise applications are designed with a microservice architecture~\cite{newman2015building}.

However, testing RESTful web services poses several challenges.
In the literature, several techniques have been proposed for automatically generating test cases in many different testing contexts.
But, as far as we know, we are aware of no technique that could automatically generate integration,  white-box tests for RESTful web services.
This kind of tests are what often engineers write during the development of their web services, using, for example, the very popular library RestAssured.

In this paper, we have proposed a technique to automatically collect white-box information from the running web services, and, then, exploit such information to generate test cases using an evolutionary algorithm.
We have implemented our novel approach in a tool prototype called \evo, written in Kotlin/Java, and ran experiments on three different web services.
Two of them  are existing open-source projects, available on GitHub.
The third was a web service provided by one of our industrial partners.
These services range from 2 to 10 thousand lines of code (existing tests included). 

Our technique was able to generate test cases which did find 38 bugs in those web services.
However, compared to the existing test cases in those projects, achieved coverage was lower.
A manual analysis of results pointed out to three different main problems: handling of string constraints, accesses to database and to other external services.

Future work will need to focus on these three main issues.
Furthermore, to achieve a wider impact in industry, it will also be important to extend our tool to also handle other popular languages in which RESTful web services are often written in, like for example JavaScript/NodeJS and C\#.
Due to a clean separation between the testing tool (written in Kotlin) and the library to collect and export white-box information (written in Java, but technically usable for any JVM language), supporting a new language is just a matter of re-implementing that library, not the whole tool.
To make the integration of different languages simpler, our library itself is designed as a RESTful web service where the coverage information is exported in JSON format.
However, code instrumentation (e.g., bytecode manipulation in the JVM) can be quite different among languages.

To learn more about \evo, visit our webpage at:
\texttt{www.evomaster.org}

\section*{Acknowledgment}
We would like to thank Andreas Bi{\o}rn-Hansen for valuable feedback, and our industrial partners for providing one of the RESTful services. 
This work is supported by the National Research Fund, Luxembourg (FNR/P10/03).



%

\bibliographystyle{IEEEtran}
\bibliography{../../papers}

\begin{thebibliography}{10}
\providecommand{\url}[1]{#1}
\csname url@samestyle\endcsname
\providecommand{\newblock}{\relax}
\providecommand{\bibinfo}[2]{#2}
\providecommand{\BIBentrySTDinterwordspacing}{\spaceskip=0pt\relax}
\providecommand{\BIBentryALTinterwordstretchfactor}{4}
\providecommand{\BIBentryALTinterwordspacing}{\spaceskip=\fontdimen2\font plus
\BIBentryALTinterwordstretchfactor\fontdimen3\font minus
  \fontdimen4\font\relax}
\providecommand{\BIBforeignlanguage}[2]{{%
\expandafter\ifx\csname l@#1\endcsname\relax
\typeout{** WARNING: IEEEtran.bst: No hyphenation pattern has been}%
\typeout{** loaded for the language `#1'. Using the pattern for}%
\typeout{** the default language instead.}%
\else
\language=\csname l@#1\endcsname
\fi
#2}}
\providecommand{\BIBdecl}{\relax}
\BIBdecl

\bibitem{newman2015building}
S.~Newman, \emph{Building Microservices}.\hskip 1em plus 0.5em minus
  0.4em\relax " O'Reilly Media, Inc.", 2015.

\bibitem{fielding2000architectural}
R.~T. Fielding, ``Architectural styles and the design of network-based software
  architectures,'' Ph.D. dissertation, University of California, Irvine, 2000.

\bibitem{canfora2009service}
G.~Canfora and M.~Di~Penta, ``{Service-oriented architectures testing: A
  survey},'' in \emph{Software Engineering}.\hskip 1em plus 0.5em minus
  0.4em\relax Springer, 2009, pp. 78--105.

\bibitem{bozkurt2013testing}
M.~Bozkurt, M.~Harman, and Y.~Hassoun, ``Testing and verification in
  service-oriented architecture: a survey,'' \emph{Software Testing,
  Verification and Reliability (STVR)}, vol.~23, no.~4, pp. 261--313, 2013.

\bibitem{GoA_TSE12}
G.~Fraser and A.~Arcuri, ``Whole test suite generation,'' \emph{IEEE
  Transactions on Software Engineering}, vol.~39, no.~2, pp. 276--291, 2013.

\bibitem{rodriguez2016rest}
C.~Rodr{\'\i}guez, M.~Baez, F.~Daniel, F.~Casati, J.~C. Trabucco, L.~Canali,
  and G.~Percannella, ``Rest apis: a large-scale analysis of compliance with
  principles and best practices,'' in \emph{International Conference on Web
  Engineering}.\hskip 1em plus 0.5em minus 0.4em\relax Springer, 2016, pp.
  21--39.

\bibitem{AIB11}
A.~Arcuri, M.~Z. Iqbal, and L.~Briand, ``Random testing: Theoretical results
  and practical implications,'' \emph{IEEE Transactions on Software Engineering
  (TSE)}, vol.~38, no.~2, pp. 258--277, 2012.

\bibitem{harman2012search}
M.~Harman, S.~A. Mansouri, and Y.~Zhang, ``Search-based software engineering:
  Trends, techniques and applications,'' \emph{ACM Computing Surveys (CSUR)},
  vol.~45, no.~1, p.~11, 2012.

\bibitem{ABHP09}
S.~Ali, L.~Briand, H.~Hemmati, and R.~Panesar-Walawege, ``{A systematic review
  of the application and empirical investigation of search-based test-case
  generation},'' \emph{IEEE Transactions on Software Engineering (TSE)},
  vol.~36, no.~6, pp. 742--762, 2010.

\bibitem{fraser2011evosuite}
G.~Fraser and A.~Arcuri, ``Evo{S}uite: automatic test suite generation for
  object-oriented software,'' in \emph{ACM Symposium on the Foundations of
  Software Engineering (FSE)}, 2011, pp. 416--419.

\bibitem{fraser2014large}
------, ``{A large-scale evaluation of automated unit test generation using
  EvoSuite},'' \emph{ACM Transactions on Software Engineering and Methodology
  (TOSEM)}, vol.~24, no.~2, p.~8, 2014.

\bibitem{canfora2006testing}
G.~Canfora and M.~Di~Penta, ``Testing services and service-centric systems:
  Challenges and opportunities,'' \emph{IT Professional}, vol.~8, no.~2, pp.
  10--17, 2006.

\bibitem{bertolino2012trends}
A.~Bertolino, G.~De~Angelis, A.~Sabetta, and A.~Polini, ``Trends and research
  issues in soa validation,'' in \emph{Performance and Dependability in Service
  Computing: Concepts, Techniques and Research Directions}.\hskip 1em plus
  0.5em minus 0.4em\relax IGI Global, 2012, pp. 98--115.

\bibitem{xu2005testing}
W.~Xu, J.~Offutt, and J.~Luo, ``Testing web services by xml perturbation,'' in
  \emph{Software Reliability Engineering, 2005. ISSRE 2005. 16th IEEE
  International Symposium on}.\hskip 1em plus 0.5em minus 0.4em\relax IEEE,
  2005, pp. 10--pp.

\bibitem{bai2005wsdl}
X.~Bai, W.~Dong, W.-T. Tsai, and Y.~Chen, ``Wsdl-based automatic test case
  generation for web services testing,'' in \emph{Service-Oriented System
  Engineering, 2005. SOSE 2005. IEEE International Workshop}.\hskip 1em plus
  0.5em minus 0.4em\relax IEEE, 2005, pp. 207--212.

\bibitem{martin2006automated}
E.~Martin, S.~Basu, and T.~Xie, ``Automated robustness testing of web
  services,'' in \emph{Proceedings of the 4th International Workshop on SOA And
  Web Services Best Practices (SOAWS 2006)}, 2006.

\bibitem{ma2008wsdl}
C.~Ma, C.~Du, T.~Zhang, F.~Hu, and X.~Cai, ``Wsdl-based automated test data
  generation for web service,'' in \emph{Computer Science and Software
  Engineering, 2008 International Conference on}, vol.~2.\hskip 1em plus 0.5em
  minus 0.4em\relax IEEE, 2008, pp. 731--737.

\bibitem{bartolini2009ws}
C.~Bartolini, A.~Bertolino, E.~Marchetti, and A.~Polini, ``Ws-taxi: A
  wsdl-based testing tool for web services,'' in \emph{Software Testing
  Verification and Validation, 2009. ICST'09. International Conference
  on}.\hskip 1em plus 0.5em minus 0.4em\relax IEEE, 2009, pp. 326--335.

\bibitem{li2016generating}
Y.~Li, Z.-a. Sun, and J.-Y. Fang, ``Generating an automated test suite by
  variable strength combinatorial testing for web services,'' \emph{CIT.
  Journal of Computing and Information Technology}, vol.~24, no.~3, pp.
  271--282, 2016.

\bibitem{bozkurt2011automatically}
M.~Bozkurt and M.~Harman, ``Automatically generating realistic test input from
  web services,'' in \emph{Service Oriented System Engineering (SOSE), 2011
  IEEE 6th International Symposium on}.\hskip 1em plus 0.5em minus 0.4em\relax
  IEEE, 2011, pp. 13--24.

\bibitem{wotawa2013fifty}
F.~Wotawa, M.~Schulz, I.~Pill, S.~Jehan, P.~Leitner, W.~Hummer, S.~Schulte,
  P.~Hoenisch, and S.~Dustdar, ``Fifty shades of grey in soa testing,'' in
  \emph{Software Testing, Verification and Validation Workshops (ICSTW), 2013
  IEEE Sixth International Conference on}.\hskip 1em plus 0.5em minus
  0.4em\relax IEEE, 2013, pp. 154--157.

\bibitem{jehan2014soa}
S.~Jehan, I.~Pill, and F.~Wotawa, ``Soa testing via random paths in bpel
  models,'' in \emph{Software Testing, Verification and Validation Workshops
  (ICSTW), 2014 IEEE Seventh International Conference on}.\hskip 1em plus 0.5em
  minus 0.4em\relax IEEE, 2014, pp. 260--263.

\bibitem{bartolini2011whitening}
C.~Bartolini, A.~Bertolino, S.~Elbaum, and E.~Marchetti, ``Bringing white-box
  testing to service oriented architectures through a service oriented
  approach,'' \emph{Journal of Systems and Software (JSS)}, vol.~84, no.~4, pp.
  655--668, 2011.

\bibitem{ye2013whitening}
C.~Ye and H.-A. Jacobsen, ``Whitening soa testing via event exposure,''
  \emph{IEEE Transactions on Software Engineering (TSE)}, vol.~39, no.~10, pp.
  1444--1465, 2013.

\bibitem{chakrabarti2009test}
S.~K. Chakrabarti and P.~Kumar, ``Test-the-rest: An approach to testing restful
  web-services,'' in \emph{Future Computing, Service Computation, Cognitive,
  Adaptive, Content, Patterns, 2009. COMPUTATIONWORLD'09. Computation
  World:}.\hskip 1em plus 0.5em minus 0.4em\relax IEEE, 2009, pp. 302--308.

\bibitem{lamela2013towards}
P.~Lamela~Seijas, H.~Li, and S.~Thompson, ``Towards property-based testing of
  restful web services,'' in \emph{Proceedings of the twelfth ACM SIGPLAN
  workshop on Erlang}.\hskip 1em plus 0.5em minus 0.4em\relax ACM, 2013, pp.
  77--78.

\bibitem{chakrabarti2010connectedness}
S.~K. Chakrabarti and R.~Rodriquez, ``Connectedness testing of restful
  web-services,'' in \emph{Proceedings of the 3rd India software engineering
  conference}.\hskip 1em plus 0.5em minus 0.4em\relax ACM, 2010, pp. 143--152.

\bibitem{pinheiro2013model}
P.~V.~P. Pinheiro, A.~T. Endo, and A.~Simao, ``Model-based testing of restful
  web services using uml protocol state machines,'' in \emph{Brazilian Workshop
  on Systematic and Automated Software Testing}, 2013.

\bibitem{fertig2015model}
T.~Fertig and P.~Braun, ``Model-driven testing of restful apis,'' in
  \emph{Proceedings of the 24th International Conference on World Wide
  Web}.\hskip 1em plus 0.5em minus 0.4em\relax ACM, 2015, pp. 1497--1502.

\bibitem{di2007search}
M.~Di~Penta, G.~Canfora, G.~Esposito, V.~Mazza, and M.~Bruno, ``Search-based
  testing of service level agreements,'' in \emph{Genetic and Evolutionary
  Computation Conference (GECCO)}.\hskip 1em plus 0.5em minus 0.4em\relax ACM,
  2007, pp. 1090--1097.

\bibitem{barr2015oracle}
E.~T. Barr, M.~Harman, P.~McMinn, M.~Shahbaz, and S.~Yoo, ``The oracle problem
  in software testing: A survey,'' \emph{IEEE Transactions on Software
  Engineering (TSE)}, vol.~41, no.~5, pp. 507--525, 2015.

\bibitem{evo1600faults2015}
G.~Fraser and A.~Arcuri, ``1600 faults in 100 projects: automatically finding
  faults while achieving high coverage with evosuite,'' \emph{Empirical
  Software Engineering (EMSE)}, vol.~20, no.~3, pp. 611--639, 2015.

\bibitem{Mcm04}
P.~McMinn, ``Search-based software test data generation: A survey,''
  \emph{Software Testing, Verification and Reliability}, vol.~14, no.~2, pp.
  105--156, 2004.

\bibitem{Kor90}
B.~Korel, ``Automated software test data generation,'' \emph{IEEE Transactions
  on Software Engineering}, pp. 870--879, 1990.

\bibitem{rojas2016detailed}
J.~M. Rojas, M.~Vivanti, A.~Arcuri, and G.~Fraser, ``A detailed investigation
  of the effectiveness of whole test suite generation,'' \emph{Empirical
  Software Engineering (EMSE)}, pp. 1--42, 2016.

\bibitem{FrA12b}
G.~Fraser and A.~Arcuri, ``Sound empirical evidence in software testing,'' in
  \emph{ACM/IEEE International Conference on Software Engineering (ICSE)},
  2012, pp. 178--188.

\bibitem{arcuri2013parameter}
A.~Arcuri and G.~Fraser, ``Parameter tuning or default values? an empirical
  investigation in search-based software engineering,'' \emph{Empirical
  Software Engineering (EMSE)}, vol.~18, no.~3, pp. 594--623, 2013.

\bibitem{Alshraideh06}
M.~Alshraideh and L.~Bottaci, ``Search-based software test data generation for
  string data using program-specific search operators,'' \emph{Software
  Testing, Verification, and Reliability}, vol.~16, no.~3, pp. 175--203, 2006.

\bibitem{rojas2016seeding}
J.~M. Rojas, G.~Fraser, and A.~Arcuri, ``Seeding strategies in search-based
  unit test generation,'' \emph{Software Testing, Verification and Reliability
  (STVR)}, 2016.

\end{thebibliography}

\end{document}